\documentclass[prl,twocolumn,showpacs,epsfig]{revtex4}
\usepackage{graphicx}
\usepackage{epsf,epstopdf,wrapfig}
\usepackage{graphics}
\usepackage{wrapfig}
\usepackage{dcolumn}
\usepackage{amssymb}
\usepackage{bm}
\usepackage{epsfig}
\usepackage{hyperref}
\usepackage{psfrag}
\usepackage{color}
\usepackage[utf8x]{inputenc}
\usepackage{ucs}
\usepackage{amsmath}
\usepackage{amsfonts}
\usepackage{amssymb}
\usepackage{graphicx}
\usepackage{dcolumn}
\usepackage{bm}
\usepackage{latexsym}
\usepackage{hyperref}
\newcommand{\bsf}[1]{\textsf{\textbf{#1}}}
\def\Jt{\tilde{J}}
\def\Dt{\tilde{\mathcal{A}}}

\def\3dots{\:\raisebox{-0.5ex}{$\stackrel{\textstyle.}{:}$}\:}
\def\beq{\begin{equation}}
\def\eeq{\end{equation}}
\def\bea{\begin{eqnarray}}
\def\eea{\end{eqnarray}}

\begin{document}

\title{Asymmetric exchange in flocks}
\author{Lokrshi Prawar Dadhichi$^1$, Rahul Chajwa$^1$, Ananyo Maitra$^2$ and Sriram Ramaswamy$^{1}$}
\altaffiliation{On leave from the Department of Physics, Indian Institute
of Science, Bangalore}\email{sriram@tifrh.res.in}
\affiliation{$^1$TIFR Centre for Interdisciplinary Sciences, Tata Institute of Fundamental Research, 21 Brundavan Colony, Osman Sagar Road, Narsingi, Hyderabad 500 075, India\\$^2$LPTMS - Universit\'e Paris Sud, 91405 Orsay Cedex, France}

\date{\today}
\begin{abstract}
As the constituents of a flock are polar, one expects a fore-aft asymmetry in their interactions. We show here that the resulting antisymmetric part of the ``exchange coupling'' between a bird and its neighbours, if large enough, destabilizes the flock through spontaneous turning of the birds. The same asymmetry also yields a natural mechanism for a difference between the speed of advection of information along the flock and the speed of the flock itself. We show that the absence of detailed balance, and not merely the breaking of Galilean invariance, is responsible for this difference. We delineate the conditions on parameters and wavenumber for the existence of the turning instability. Lastly we present an alternative perspective based on flow-alignment effects in an active liquid crystal with turning inertia in contact with a momentum sink. 	
\end{abstract}
\maketitle

In the classic models of flocking \cite{vicsek,tonertu} and much of the later literature \cite{otherflock} each agent (hereafter termed a bird) is assumed to adjust its direction of motion instantaneously to the mean of its neighbours including itself, plus a random error. It has recently become clear \cite{ISM} that, on time- and length-scales relevant to observations on real bird flocks, the dynamics of this reorientation must be explicitly taken into account, via an explicit classical spin angular momentum on which the aligning interaction acts as a torque. This inertial effect was shown \cite{ISM,silent,XY} to give rise, on intermediate length scales \cite{Ram_maz}, to turning waves reminiscent of those predicted for inertial flocks in fluids \cite{simhaSR} or rotor 
lattices \cite{chailub}. On the longest scales, where damping by the ambient medium overcomes inertia, the dynamics is effectively described by the Toner-Tu \cite{tonertu} equations. 

\begin{figure}
  \includegraphics[width=3 truecm]{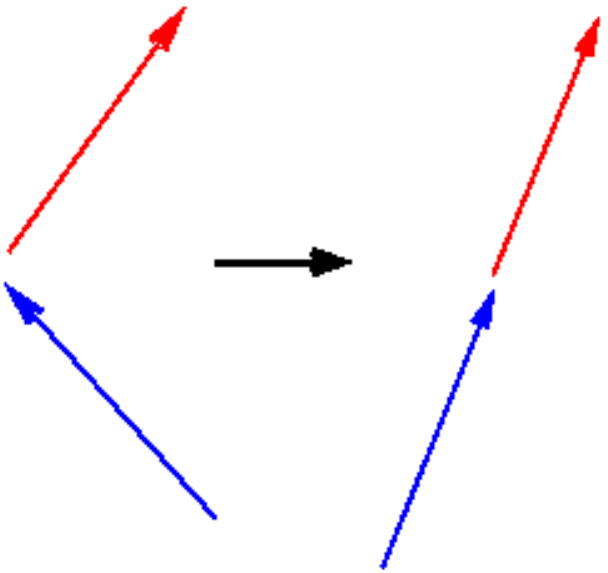}
  \caption{A pair of fore-aft-separated birds with misaligned velocities; for reasons of vision or airflow, the bird in front is slower to adjust its orientation. This asymmetric ``exchange coupling'' leads to a net rotation of the overall alignment of the birds.}
  \label{fig:chiral_rotate}
\end{figure}

However, in \cite{ISM,silent} the aligning field was implicitly taken to arise from an effective Hamiltonian, so the interactions were perfectly mutual. A pair of birds exerted opposing torques of equal magnitude on each other, conserving spin angular momentum. This is in principle unduly restrictive: terms not governed by an energy function are permitted in systems out of thermal equilibrium, and the dynamics takes place in contact with an ambient medium with which the birds can exchange both angular and linear momentum. Indeed, self-propulsion consists precisely in drawing linear momentum from the ambient medium, with directional bias determined by the structural polarity of the bird. Despite the possibility of drawing \textit{angular} momentum from the ambient air, self-propelling activity will not lead to a net persistent rotational motion as a bird can reasonably be considered achiral on average. In the flocking models we consider, birds are individually achiral and carry only a position and a vectorial orientation. Transient chirality and, hence, spontaneous rotation can arise only in the relative arrangements of two or more birds.
Consider a pair of birds flying one ahead of the other. The basic assumption of flocking models is that if the velocity vectors of the birds depart slightly from being parallel, an aligning torque arises. Each bird tries to rotate its flying orientation to match that of its neighbour, but one expects in general that the aligning response of the leading bird to the trailing bird should be different from that of the trailer to the leader (Fig. \ref{fig:chiral_rotate}). Such antisymmetry of information transfer or response could arise, inter alia, from vision \cite{birdvisionwiki} or airflow \cite{jeb2011}. Our focus is distinct from that of ref. \cite{trigger} in which pairwise antisymmetry, distributed statistically across a collection of birds, gives rise to an ultra-rapid response and relaxation. In both works, however, motility and antisymmetric aligning torques enter as two distinct manifestations of the nonequilibrium nature of the system.

Here we study the consequence of this physically natural asymmetry of interaction, within the hydrodynamic description \cite{silent} of flocks with turning inertia \cite{ISM}. 
The interaction between birds flying precisely side-by-side is taken to be symmetric. Our main results are: (i) If the antisymmetric part $\mathcal{A}$ of the aligning interaction between fore-aft separated birds exceeds a threshold $\mathcal{A}_c \propto \sqrt{J}$ where $J$ is the symmetric part, uniform flocks undergo a spontaneous buckling instability with wavevector ${\bf k}$ along the mean direction of alignment of the flock, which we denote $\vert \vert$. (ii) A crossover lengthscale $\xi \propto \sqrt{J (1+\mathcal{A}/\mathcal{A}_c)}$ separates the dynamics into two regimes. 
For $k_{||} \xi \ll 1$ the instability growth rate is diffusive, $\sim (\mathcal{A} - \mathcal{A}_c) k_{||}^2$, crossing over to $(\mathcal{A} - \mathcal{A}_c) \xi^{-2}$ for $k_{||} \xi \gg 1$. (iii) For $\mathcal{A} < \mathcal{A}_c$, where the flock is stable, the small-$k$ dynamics is of the Toner-Tu type, with the coefficient of the advective term shifted by a contribution proportional to $\mathcal{A}$. At large $k$ the turning waves of refs. \cite{ISM,silent} are recovered. Antisymmetric exchange thus plays a dual role, providing a natural mechanism for a difference between the information-transfer speed and the flock speed \textit{and}, if large enough, destabilizing the flock. (iv) Lastly, we show that antisymmetric exchange emerges from the classic flow-alignment term \cite{flowalign} if motility is introduced into polar liquid crystal hydrodynamics \cite{kung}, with rotational inertia and without momentum conservation. 

Before showing how we arrived at these results, a remark is in order regarding advective effects in flocking models. In the Toner-Tu \cite{tonertu} equation the velocity field does not advect itself at the same rate as it advects the density. While it is tempting to attribute this asymmetry simply to the absence of Galilean invariance in the theory, both Galilean invariance and detailed balance have to be absent for this effect to emerge \cite{ananyo_thesis}. However, when constructing Toner-Tu hydrodynamics from inertial spin hydrodynamics \cite{ISM,silent} this difference in advection velocities has to be put in by hand. As we shall see below, the simplest microscopic form of the Inertial Spin Model (ISM) \cite{ISM} leads to equal coefficients for advection in the Toner-Tu density and velocity equations. Our result (iii) above thus clarifies that an antisymmetric exchange coupling in the ISM is essential in order for the advective term in the resultant Toner-Tu equation to have a coefficient different from unity. With this remark in hand we turn to showing how we obtained our results. 

Consider a flock in the $xy$ plane. Let the $\alpha$th bird have velocity ${\bf  
v}_{\alpha} \equiv v_0 \hat{\bf v}_{\alpha}$ with fixed magnitude $v_0$ and 
classical ``spin'' angular momentum ${\bf 
s}_{\alpha}$, about its centre of mass, along $\hat{\bf z}$. Let us describe the aligning 
interaction of neighbouring birds as a torque  
\beq
\label{torque}
\dot{\bf s}_{\alpha} = 
\sum_{\beta}J_{\alpha \beta}\hat{\bf v}_{\alpha} \times \hat{\bf v}_{\beta}
\eeq
due to birds $\beta$ neighbouring $\alpha$, which rotates the direction of the velocities:  
\beq
\label{rotation}
\dot{\bf v}_{\alpha} = {{\bf s}_{\alpha} \over \chi} \times {\bf v}_{\alpha}, 
\eeq
where $\chi$ is a rotational inertia \cite{ISM,silent}. As remarked above we must allow for processes 
that do not conserve angular momentum \cite{air}. Specifically, we must 
allow for the possibility that the 
coupling $J_{\alpha \beta}$ is non-symmetric. The result is that the rate at 
which bird $\alpha$ turns to align with $\beta$ differs from that at which 
$\beta$ turns to align with $\alpha$. Such ``antisymmetric exchange'' 
\cite{jayajit,DM} violates angular momentum conservation while preserving rotation 
invariance, because the aligning field does not arise from an energy 
function \cite{behav}, and is under no obligation to do so. Making 
the physically reasonable approximation that the interactions of birds are 
left-right symmetric, we see that $J_{\alpha \beta}$ should have an antisymmetric 
part when birds $\alpha$ and $\beta$ are one behind the other -- possibly 
through polar asymmetries of airflow or vision -- and not when they are 
side-by-side. Of course, even the symmetric part of $J_{\alpha \beta}$ 
should in general be anisotropic and should thus differ for longitudinal and lateral 
neighbours. However, this latter asymmetry is not of much consequence for the issues considered here; at least within a linearized theory it can be removed by anisotropic rescaling of coordinates.   

For a one-dimensional chain of birds oriented in the direction of mean 
motion, we therefore allow for nearest neighbour couplings 
\cite{nonmetric} 
\begin{equation}
J_{\alpha, \alpha \pm 1} = \Jt \pm \Dt, 
\end{equation}
leading via the torque (\ref{torque}) to the equation of motion 
\begin{equation}
\label{particleeq}
\dot{\mathbf{s}}_{\alpha} = {\Jt \over v_0^2} \mathbf{v}_{\alpha} \times 
(\mathbf{v}_{\alpha + 1} + \mathbf{v}_{\alpha - 1}) +{\Dt \over v_0^2} 
\mathbf{v}_{\alpha} \times(\mathbf{v}_{\alpha + 
1} - \mathbf{v}_{\alpha - 1})  - \frac{\eta}{\chi} 
\mathbf{s}_{\alpha}
\end{equation}
where $\eta$ is friction with the ambient medium. Going beyond a one-dimensional array with fixed neighbour assignments, fore-aft asymmetry must be defined with respect to the orientation (and hence velocity) vector of a bird. Let us pass from a Lagrangian picture tied to individual birds to a continuum Eulerian description referred to points in space, with number density $\rho({\bf x},t)$, velocity ${\bf v}({\bf x},t)$ and spin angular momentum density ${\bf s}({\bf x},t)$ fields as functions of position ${\bf x}$ and time $t$. Introducing a potential 
\beq
\label{Landaupotl}
U=\int d^dr\left[-{\alpha \over 2}\mathbf{v}\cdot\mathbf{v}+ {\beta \over 4}
(\mathbf{v}\cdot\mathbf{v})^2\right]
\eeq
favouring a local speed $v_0 = \sqrt{\alpha / \beta}$, and a 
pressure $P(\rho)$, we obtain the hydrodynamic equations 
\begin{equation}
\label{velpde}
D_t\mathbf{v} = \frac{1}{\chi}\mathbf{s}\times\mathbf{v} - {1 \over \rho}\nabla P - 
\frac{\delta U}{\delta\mathbf{v}}, 
\end{equation}
\begin{equation}
\label{spinpde}
D_t\mathbf{s} = {J \over v_0^2} \mathbf{v}\times\nabla^2\mathbf{v}  
+{\mathcal{A} \over v_0^3} \mathbf{v}\times({\bf v} \cdot \nabla \mathbf{v}) - 
\frac{\eta}{\chi} \mathbf{s}
\end{equation}
and
\begin{equation}
\label{conteq}     
     \partial_t\rho = -\nabla.(\rho\mathbf{v}), 
\end{equation}
where $D_t = 
\partial_t + \mathbf{v} \cdot \nabla$ is the material derivative, without the possibility at this stage of an arbitrary advection coefficient, and the antisymmetric coupling $\mathcal{A}$ acts between fore-aft neighbours as defined {\em locally} by ${\bf v}$. Note that in replacing the differences in (\ref{particleeq}) by the spatial derivatives in (\ref{spinpde}) we have implicitly introduced a mean interbird spacing $a$, in terms of which $J = a^2\Jt$ and $\mathcal{A} = 2a\Dt$. The ratio $J/\mathcal{A}$ thus has units of length. Equations (\ref{velpde}) and (\ref{conteq}) are as in the original inertial spin model \cite{ISM}, but (\ref{spinpde}) has the new term $\mathcal{A} 
\hat{\bf v}\times\partial_\parallel \hat{\bf v}$ from the antisymmetric exchange 
coupling, where $\partial_{\parallel} = \hat{\bf v} \cdot \nabla$ 
%

Eq. \eqref{spinpde} tells us that $\mathbf{s}$ is a fast variable, relaxing on non-hydrodynamic timescales to a value determined by $\mathbf{v}$ and $\rho$. Implementing this prescription \textit{to leading order in }$1/\eta$, one can solve \eqref{spinpde} by disregarding the time derivative and insert the resulting value for $\mathbf{s}$ in (\ref{velpde}), yielding 
\begin{equation}
\label{tonertulimit}
\partial_t {\bf v}+ (1 + {\mathcal{A} \over \eta}\bsf{T}) \mathbf{v} \cdot 
\nabla \mathbf{v} = {J \over \eta}  
\bsf{T}\nabla^2 \mathbf{v} -{\delta U \over \delta {\bf v}} - {1 \over \rho} \nabla P, 
\end{equation} 
where $\bsf{T} = {\bf I} - \hat{\bf v} \hat{\bf v}$ is the projector transverse to ${\bf v}$, and the continuity equation (\ref{conteq}) as before.   
We thus recover the Toner-Tu \cite{tonertu} equations apart from innocuous projectors transverse 
to ${\bf v}$ which arise because of the length-preserving nature of the torque (\ref{torque}). Interestingly, the antisymmetry parameter $\mathcal{A}$ provides a natural mechanism for a coefficient different from unity for the advection term in \eqref{tonertulimit}, a key feature of the Toner-Tu \cite{tonertu} formulation. However, losing Galilean invariance alone cannot generate such a term: the absence of detailed  balance, as signalled by $\mathcal{A}$, is crucial \cite{ananyo_thesis}. In an \textit{equilibrium} fluid moving in contact with a momentum sink the coefficients in the momentum and density equations must agree. Note that $\mathcal{A}$ leads to a fore-aft asymmetry in the local transfer of orientational information. It is therefore natural that it should give rise to a disturbance speed different from the flock speed. 

We will see below that the above leading-order treatment is a little misleading, as it fails to capture some important physics at order $\nabla^2$ in (\ref{tonertulimit}). To this end let us study the coupled dynamics of (\ref{velpde}) - (\ref{conteq}) in more detail, 
linearizing about an ordered steadily moving state: $\mathbf{v}={\bf v}_0 +\delta\mathbf{v}$, $\rho=\rho_0 
+\delta\rho$, ${\bf s} = {\bf 0} + s\hat{\bf z}$. Defining $\sigma = 
P'(\rho_0)$ and directions $\Vert$ and $\perp$ along and transverse 
to ${\bf v}_0$ and working in a frame moving with same velocity as the mean 
velocity of flock we get 
        \begin{equation}
        \label{vperpeq}
        \partial_t \delta v_\bot = -\sigma\partial_\bot \delta\rho + 
\frac{v_0}{\chi}s
        \end{equation}
        \begin{equation}
        \partial_t s = {J\over v_0}\nabla^2\delta v_\perp + {\mathcal{A} 
\over v_0} \partial_\parallel\delta v_\perp - \frac{\eta}{\chi}s
        \end{equation}
      \begin{equation}
      \partial_t\delta\rho = -\rho_0\partial_\perp\delta v_\perp. 
      \end{equation}
The coupling of orientation to spin angular momentum through $\mathcal{A}$ 
enters only the modes with $k_{\Vert} \neq 0$, while orientation and density 
couple through splay ($k_{\perp}$). These modes can thus be separated by 
looking at ${\bf k}$ along or transverse to $\Vert$. The general characteristic 
equation relating frequency $\omega$ to wavevector ${\bf k} = 
(k_{\Vert}, k_{\perp})$ reads  
     \begin{equation}
     \label{charac}
     \omega^3 + i \frac{\eta}{\chi}\omega^2 + i 
\frac{\mathcal{A}}{\chi} k_\parallel\omega - \frac{J}{\chi} 
k^2\omega - \sigma\rho_0 k^2_ \perp \omega - i 
\sigma\rho_0\frac{\eta}{\chi}k^2_\perp = 0.
     \end{equation}
To highlight the physics of antisymmetric exchange we look at modes with 
$k_{\perp}=0$, for which 
     \begin{equation}
     \omega^2 + i \frac{\eta}{\chi}\omega + i 
\frac{\mathcal{A}}{\chi} k_\parallel - \frac{J}{\chi} k_\parallel^2 = 
0
     \end{equation}
which yields mode frequencies 
     \begin{equation}
     \label{bendfreq}
     \omega = -i \frac{\eta}{2\chi} \pm i \frac{\eta}{2\chi}\sqrt{1 + 
\frac{4\chi^2}{\eta^2}\left(i \frac{\mathcal{A}}{\chi}k_\parallel - 
\frac{J}{\chi}k^2_\parallel\right)}. 
     \end{equation}
The lower sign corresponds to the fast decay of $s$, at a rate $\eta/\chi$ in the limit of zero wavenumber. $s$ nonetheless affects the dynamics of $v_{\perp}$ at nonzero wavenumber.  
For small $k_{||}$, i.e., in the Toner-Tu regime, the eigenfrequencies are 
\begin{eqnarray}\label{smallk}
&&\hskip-1.2cm \omega  =  \left \{ \begin{array}{c}
    -{\mathcal{A} \over \eta}k_{\vert\vert} - i  \left(J - \mathcal{A}^2 {\chi \over \eta^2} \right){k_{\vert\vert}^2 \over \eta},             \\\\
      {\mathcal{A} \over \eta} k_{\vert\vert} - i {\eta \over \chi}        
           \end{array} \right. 
\end{eqnarray}
so that the correct Toner-Tu limit is one in which $J$ in (\ref{tonertulimit}) is replaced by $J - \mathcal{A}^2 \chi / \eta^2$. For large $k_{||}$, that is, in the turning-wave regime, the modes cross over to 
\beq
\label{largek}
\omega \simeq \pm \sqrt{J \over \chi} k_{\vert \vert} - {i \over 2 } \left({\eta \over \chi} \pm {\mathcal{A} \over \sqrt{\chi J}}\right).  
\eeq
In either limit and for all wavenumbers in between we see that the uniformly moving flock suffers a buckling instability for $|\mathcal{A}|$ larger than a threshold 
\begin{equation}
\label{instabcond}
\mathcal{A}_c \equiv \eta \sqrt{J / \chi}.  
\end{equation}
The instability is diffusive for small $k_{||}$ and wavenumber-independent for large $k_{||}$. For $\mathcal{A} > 0$, which corresponds to the case where a bird responds more to the bird ahead of it than to the bird behind, we see from (\ref{smallk}) and (\ref{largek}) that the unstable disturbance travels towards the rear of the flock, which is physically reasonable. The propagation speed, though, is different in the two regimes. At small $k$ it is set by the kinematic-wave speed along the flock, determined by $\mathcal{A}$ while at large $k$ the relevant speed $\sqrt{J/\chi}$ is that of the turning waves, as in \cite{ISM,silent}. It is readily seen that the crossover between the growth rates in (\ref{smallk}) and (\ref{largek}) takes place at $k_{||} \xi \sim 1$ where  
\beq
\label{xidef}
\xi^2 = {J \chi \over \eta^2} \left(1 + {\mathcal{A} \over \eta} \sqrt{\chi \over J}\right). 
\eeq 
We have thus derived the results (i) - (iii) advertised at the beginning of this work. 
We note in passing that the effect of the antisymmetric exchange $\mathcal{A}$ can be made especially prominent by considering wavenumbers   
\beq
\label{kcondition}
      \frac{\eta^2}{4\chi \mathcal{A} 
} \ll k_\parallel \ll \frac{\mathcal{A}}{J}, 
\eeq 
for which \eqref{bendfreq} simplifies to 
     \begin{equation}
\label{asym_instab}
\omega = \pm {1-i \over \sqrt{2}} 
\sqrt{\frac{\mathcal{A}}{\chi} k_\parallel}, 
     \end{equation}
i.e., the crossover from diffusive to saturated is through a $k_{||}^{1/2}$ dependence over a regime whose width grows with $\mathcal{A}$. The condition (\ref{kcondition}) on wavenumber can of course obtain only if $\mathcal{A} \gg \mathcal{A}_c$ as defined in \eqref{instabcond}.

To go beyond this linear stability analysis we study equations (\ref{torque}) and (\ref{rotation}) numerically in the presence of spin-damping $-(\eta/\chi){\bf s}_\alpha$ because of the ambient medium. We restrict our attention to one-dimensional strands of birds, with initial velocities (i.e., orientation vectors) globally aligned either (a) along or (b) perpendicular to the strand, plus a perturbation. For (b) we consider cases where the perturbation in ${\bf s}_{\alpha}$ is zero. Consistent with the linear stability analysis, we find unstable growth of small perturbations when $\mathcal{A}$ is large enough in case (a); see \cite{supplement}, video 1. Interestingly, even the linearly stable case (b) displays a nonlinear instability at large $\mathcal{A}$ when the initial perturbation in ${\bf v}_{\alpha}$ is large enough; see \cite{supplement}, video 2.

We further note that the instability of the longitudinal diffusivity is governed by the same term that produces the nonlinearity responsible for stabilising long-range order in two-dimensions in the Toner-Tu theory. Thus, increasing $\mathcal{A}$ \textit{ceteris paribus} always destabilizes the flock. Nonetheless, it is perfectly possible to have stable flocks at arbitrarily large $\mathcal{A}$ and, thus, arbitrarily high speed of disturbance propagation in the frame of the flock, provided the diffusivity is kept positive by increasing $J$ as well. 

We turn to result (iv) promised at the start of this article. In the discussion so far, motility (the fact that the polar order parameter ${\bf v}$ is a velocity) and antisymmetric exchange [the $\mathcal{A}\mathbf{v}\times(\mathbf{v}\cdot\nabla\mathbf{v})$ term in the spin equations] entered as independent nonequilibrium effects. We now show, within polar liquid crystal hydrodynamics \cite{kung}, modified to include rotational inertia and a momentum sink, that motility leads to antisymmetric exchange. We begin with the coupled hydrodynamic equations for \textit{distinct} velocity and orientation fields $\mathbf{u}$ and $\mathbf{v}$ and spin angular momentum density field $\mathbf{s}$, together with continuity $\partial_t \rho + \nabla \cdot (\rho \mathbf{u}) = 0$ for the density field $\rho$. Note that in the process we introduce several phenomenological parameters whose meaning is clear from context.
\begin{equation}
\label{ppde}
D_t\mathbf{v}=\frac{1}{\chi}\mathbf{s}\times\mathbf{v}-\Gamma\mathbf{h}-\frac{\Gamma'_u}{|v|^2}\mathbf{v}(\mathbf{v}\cdot\mathbf{u})
\end{equation}
\begin{equation}
\label{spde}
D_t\mathbf{s}=\mathbf{v}\times\mathbf{h}+\Gamma_\omega\boldsymbol{\omega}-\frac{\eta}{\chi}\mathbf{s}-\Gamma_D\mathbf{v}\times(\mathbf{v}\cdot\bsf{D})-\Gamma_u(\mathbf{v}\times\mathbf{u}),
\end{equation}
where $\mathbf{h}$ is the molecular field conjugate to $\mathbf{v}$, $\bsf{D} =(1/2) [\nabla \mathbf{u} + (\nabla \mathbf{u})^T]/2$ is the symmetric part of the velocity gradient, $\boldsymbol{\omega}=(\nabla\times\mathbf{u})/2$, and
\begin{equation}
\label{vpde}
\rho(\partial_t + \mathbf{u} \cdot \nabla)\mathbf{u} + \Gamma \mathbf{u} =\zeta \mathbf{v} + ....
\end{equation}
where $\Gamma$ in (\ref{vpde}) denotes damping by a substrate and the ellipsis denotes pressure and viscous terms. Note that \eqref{ppde} and \eqref{spde} are exactly the equations that one would have obtained for a uniaxial liquid crystal on a substrate. That the system is intrinsically nonequilibrium enters only at one point: the forcing $\zeta \mathbf{v}$ in Eq. (\ref{vpde}) for the velocity and the polarisation. On long time and length scales (\ref{vpde}) reduces to $\mathbf{u} =\zeta \mathbf{v}/\Gamma$, allowing us to eliminate $\mathbf{u}$ in favour of $\mathbf{v}$ in (\ref{ppde}) and (\ref{spde}), the final term in \eqref{spde} turns into the term that was argued to arise because of antisymmetric exchange, with $\mathcal{A} =(\Gamma_D/2)(\zeta/\Gamma)v_0^3$. In fact, \eqref{ppde} and \eqref{spde} turn into \eqref{velpde} and \eqref{spinpde} upon rescaling $(\zeta/\Gamma){\bf v}\to{\bf v}$, with $\mathcal{A} =(\Gamma_D/2)(\Gamma/\zeta)^2v_0^3$.
We refer the interested readers to \cite{supplement} for an explicit expression of the dynamical equation (including terms at next order in gradients) that is obtained by this method. The $\mathbf{v}\times(\mathbf{v}\cdot\bsf{D})$ term describes liquid crystal flow alignment \cite{flowalign}. Thus (see \cite{RMP} for a brief discussion), flow alignment together with an active velocity in the direction of polarisation leads to a non-unit coefficient for the advective non-linearity. This implies that even if a flock does not have microscopic antisymmetric interactions, an effective antisymmetric exchange is always generated dynamically. Therefore, the theory developed in this paper remains valid even if there is no intrinsic asymmetry in the interactions between two birds.

In summary: Asymmetric exchange can't be ruled out and should therefore be part 
of the consideration in any modelling of flocks. If observations rule it out, 
it is presumably because it was selected against by evolution. If this term is 
large enough it changes qualitatively the dynamics in the ``turning inertia'' 
regime, giving rise to a spontaneous turning of the flock in either direction 
at intermediate length scales and to an ultimate disruption of the flock. At 
long length scales it gives a natural mechanism for the advective term to have 
a coefficient different from unity. It should be noted that this term, although 
rotation-invariant, is not angular-momentum-conserving and therefore can arise only if 
the birds are in contact with an ambient medium. The antisymmetry $\mathcal{A}$ and the 
air-damping $\eta$ are not entirely independent, so it remains to be seen how 
easy it is for the condition $\mathcal{A} > \mathcal{A}_c$ to be met. 

We thank A Cavagna, I Giardina, T Mora and N Rana for discussions, and all the authors of ref. \cite{trigger} for sharing their preliminary draft. SR acknowledges support from a J C Bose Fellowship.

\onecolumngrid\newpage\twocolumngrid
\setcounter{equation}{0}
\begin{widetext}

\section{Asymmetric exchange in flocks: Supplementary Material}
In this supplementary material we display the full coupled equations of density, polarisation, spin angular momentum and velocity. We then eliminate the velocity and obtain closed equations in terms of the other three fields.

\section{Dynamical equations for polar active particles with spin angular momentum}

The fields for which we will construct the dynamical equations are the density $\rho({\bf x},t)$, the polarisation ${\bf v}({\bf x}, t)$, the velocity ${\bf u}({\bf x}, t)$, and the spin angular momentum ${\bf s}({\bf x}, t)$ \cite{Stark}. 

\begin{equation}
\label{denseqn}
D_t\rho=-\rho\nabla\cdot{\bf u},
\end{equation}
\begin{equation}
\label{poleqn}
D_t{\bf v}=\frac{1}{\chi}{\bf s}\times {\bf v}-\Gamma_v{\bf h}+\lambda_1\nabla\times{\bf s}-\frac{\Gamma'_u}{|v|^2}{\bf v}({\bf v}\cdot{\bf u})+\frac{\lambda_v}{|v|^2}{\bf v}({\bf vv}:\bsf{D}),
\end{equation}
where the final two terms imply that the magnitude of polarisation changes with imposed flow, 
\begin{equation}
\label{spneqn}
D_t{\bf s}={\bf v}\times{\bf h}-\Gamma_\omega\left(\frac{{\bf s}}{\chi}-\boldsymbol{\omega}\right)+\Gamma_D{\bf v}\times({\bf v}\cdot\bsf{D})+\Gamma_D'{\bf v}\times(\nabla\cdot\bsf{D})-\lambda_1\chi\nabla\times{\bf h}-\frac{\eta-\Gamma_\omega}{\chi}{\bf s}-\Gamma_u({\bf v}\times{\bf u}),
\end{equation}
where ${\bf h}$ is the molecular field corresponding to ${\bf v}$, $D_{ij}=(\partial_iu_j+\partial_ju_i)/2$ and $\boldsymbol{\omega}=(\nabla\times{\bf u})/2$, and 
\begin{equation}
\rho D_t {\bf u}=-\Gamma{\bf u}+\zeta{\bf v}-\rho\nabla\frac{\delta F}{\delta\rho}+\nabla\cdot\boldsymbol{\sigma}
\end{equation}
with $F$ being the standard Landau-de Gennes free energy for polar liquid crystals \cite{kung} and the stress $\boldsymbol{\sigma}$ being
\begin{multline}
\boldsymbol{\sigma}=\frac{\Gamma_D}{2}[{\bf v}({\bf v}\cdot\bsf{D})]^A+\frac{\Gamma'_D}{2}[{\bf v}(\nabla\cdot\bsf{D})]^A+\frac{\Gamma_\omega}{2}\boldsymbol{\epsilon}\cdot\left(\frac{{\bf s}}{\chi}-\boldsymbol{\omega}\right)\\+\frac{\Gamma_D}{2}\left[{\bf v}\left({\bf v}\times\left(\frac{{\bf s}}{\chi}-\boldsymbol{\omega}\right)\right)\right]^S+\frac{\Gamma'_D}{2}\left[{\bf v}\left(\nabla\times\left(\frac{{\bf s}}{\chi}-\boldsymbol{\omega}\right)\right)\right]^S+\frac{\lambda_v}{2}[{\bf vv}({\bf v}\cdot{\bf h})], 
\end{multline}
where $\boldsymbol{\epsilon}\equiv\epsilon_{ijk}$ is the Levi-Civita symbol, and $\bsf{C}^A$ and $\bsf{C}^S$ denotes the anti-symmetric and symmetric parts of the tensor $\bsf{C}$ respectively. The free-energy is given by
\begin{equation}
F=\int d^d{\bf r}\left[-\frac{\alpha}{2}{\bf v}\cdot{\bf v}+\frac{\beta}{4}({\bf v}\cdot{\bf v})^2+\frac{K}{2}(\nabla{\bf v})^2+\rho\nabla\cdot{\bf v}\right]
\end{equation}

Note that these equations reduce to those for equilibrium polar liquid crystals if $\zeta=0$. If the velocity equation is taken to be overdamped while the spin angular momentum remains underdamped, we obtain the equations displayed in the main text. To the lowest order in gradients,
\begin{equation}
{\bf u}=v_p{\bf v}
\end{equation}
with $v_p=\zeta/\Gamma$. Replacing this in \eqref{denseqn}, \eqref{poleqn} and \eqref{spneqn}, we obtain
\begin{equation}
\partial_t\rho=-\nabla\cdot(v_p\rho{\bf v})
\end{equation}
\begin{equation}
\partial_t{\bf v}+v_p{\bf v}\cdot\nabla{\bf v}={\bf v}\times{\bf h}+\lambda_1\nabla\times{\bf s}-\Gamma'_uv_p{\bf v}+\frac{\lambda_vv_p}{|v|^2}{\bf v}({\bf v}\cdot\nabla|v|^2)
\end{equation}
\begin{equation}
\partial_t{\bf s}+v_p{\bf v}\cdot\nabla{\bf s}={\bf v}\times{\bf h}+\frac{\Gamma_\omega}{2} v_p\nabla\times{\bf v}+\frac{\Gamma_Dv_p}{2}{\bf v}\times\left({\bf v}\cdot\nabla{\bf v}+\frac{1}{2}\nabla|v|^2\right)+\frac{\Gamma'_Dv_p}{2}{\bf v}\times[\nabla^2{\bf v}+\nabla(\nabla\cdot{\bf v})]-\lambda_1\chi\nabla\times{\bf h}-\frac{\eta}{\chi}{\bf s}
\end{equation}
This system of equations constitute the generalised ISM model. If we now rescale $v_p{\bf v}\to {\bf v}$, we obtain exactly equation $(6)$ and $(7)$ of the main text (to lowest order in gradients).
\begin{equation}
\partial_t\rho=-\nabla\cdot(\rho{\bf v})
\end{equation}
\begin{equation}
\partial_t{\bf v}+{\bf v}\cdot\nabla{\bf v}=\frac{1}{\chi}{\bf s}\times{\bf v}-\Gamma_v{\bf h}+\lambda_1v_p\nabla\times{\bf s}-\Gamma'_uv_p{\bf v}+\frac{\lambda_v}{|v|^2}{\bf v}({\bf v}\cdot\nabla|v|^2),
\end{equation}
\begin{equation}
\label{sppde}
\partial_t{\bf s}+{\bf v}\cdot\nabla{\bf s}=\frac{1}{v_p}{\bf v}\times{\bf h}+\frac{\Gamma_\omega}{2} \nabla\times{\bf v}+\frac{\Gamma_D}{2v_p^2}{\bf v}\times\left({\bf v}\cdot\nabla{\bf v}+\frac{1}{2}\nabla|v|^2\right)+\frac{\Gamma'_D}{2v_p}{\bf v}\times[\nabla^2{\bf v}+\nabla(\nabla\cdot{\bf v})]-\lambda_1\chi\nabla\times{\bf h}-\frac{\eta}{\chi}{\bf s}
\end{equation}
This shows that $J$ receives contibutions from both the molecular field and a kinetic term (first and fourth terms on the R.H.S. of \eqref{sppde} respectively). We also see that the equation for ${\bf v}$ that one can derive by integrating out ${\bf s}$ to leading order in $1/\eta$ has three independent coefficients for the three nonlinearities with one gradient and two ${\bf v}$ -- ${\bf v}\cdot\nabla{\bf v}$, ${\bf v}\nabla\cdot{\bf v}$ and $\nabla|v|^2$ -- just as in the Toner-Tu equations. While in the traditional ISM or Toner-Tu theory, values of the nonlinearities are argued to be arbitrary purely from symmetry, our derivation here clarifies the role of activity and adsorption.
This completes our discussion of the dynamical equations of the ISM.

\section{Description of supplementary videos}
We numerically solve the following equations:
\beq
\label{torquesupp}
\dot{\bf s}_{\alpha} = 
\sum_{\beta}J_{\alpha \beta}\hat{\bf v}_{\alpha} \times \hat{\bf v}_{\beta}-\frac{\eta}{\chi}{\bf s}_\alpha
\eeq
 
\beq
\label{rotationsupp}
\dot{\bf v}_{\alpha} = {{\bf s}_{\alpha} \over \chi} \times {\bf v}_{\alpha}, 
\eeq

where the exchange interaction $J_{\alpha\beta}$ is
\begin{equation}
J_{\alpha \beta} = \Jt - \Dt  \hat{{\bf R}}_{\alpha \beta} \cdot \hat{\mathbf{v}}_{\alpha},
\end{equation}
where $\hat{{\bf R}}_{\alpha \beta}$ is the unit vector pointing from the $\beta th$ to the $\alpha th$ bird and $\hat{{\bf v}}_\alpha$ is the velocity of the $\alpha th$ bird. The summation in \eqref{torquesupp} runs over all birds within a radius $r_0$ of the $\alpha th$ one. In our calculations, $r_0=2$ i.e. the interaction range is twice the initial separation between the birds. We perform our numerical calculations for the following values of $\Jt$, $\Dt$, $\eta$  and $\chi$:

Video 1: $\Jt = 20$, $\Dt = 10$ , $\eta= 5$ and $\chi = 10$

Video 2: $\Jt = 20$, $\Dt = 10$ , $\eta= 1$ and $\chi = 10$.

The supplementary video 1 shows the linear instability of a one-dimensional flock with initial velocities of the birds oriented along the strand. The supplementary video 2 shows a \textit{nonlinear} instability of a one-dimensional flock with initial velocities oriented \textit{perpendicular} to the strand.

\end{widetext}


\begin{thebibliography}{99}
\bibitem{behav} It should be kept in mind also that the torque is 
behavioural, not mechanical, and that the birds could conceivably switch 
properties such as the antisymmetry on or off according to circumstances.
\bibitem{air} The dynamics of the air, which does not appear explicitly in our 
treatment, can be taken to balance the books in this regard, as with linear 
momentum in standard ``dry'' flocking models \cite{RMP}.
\end{thebibliography}
\end{document}